\documentclass[12pt]{article}

\input{tcilatex}
\begin{document}

\author{J.C. Kimball$^1$ and Keeyung Lee$^{1,2}$ \\
$^1$Physics Department, University at Albany\\
Albany, NY 12222\\
$^2$Physics Department, Inha University, Inchon, Korea}
\date{}
\title{A lattice model exhibiting radiation-induced anomalous conductivity}
\date{}
\maketitle

\begin{abstract}
A lattice-based model exhibits an unusual conductivity when it is subjected
to both a static magnetic field and electromagnetic radiation. This
conductivity anomaly may explain some aspects of the recently observed
``zero-resistance states''.


\end{abstract}

Electrons in large magnetic fields, specially in systems of reduced
dimension at low temperatures, exhibit a wide variety of interesting
phenomena. Recently, ``zero-resistance states'' have been reported \cite
{Mani},\cite{Zudov}. The unusual conductivity of these states is observed in
two-dimensional electron systems in magnetic fields of less than one Tesla,
temperatures around 1K, and in microwave fields with frequencies slightly
larger than integral multiples of the cyclotron resonance frequency.
Prominent among the proposed zero-resistance-state mechanisms are theories
which combine excitation by the microwaves with impurity scattering to
promote an electron to an ``uphill'' (opposing the electric force) excited
Landau level \cite{Durst}. The resulting negative microwave-induced current
may explain the observed zero-resistance states \cite{Andre}.

We describe here a quasi-one-dimensional lattice model which can produce a
microwave-induced negative contribution to the total current. This
counter-flowing current is obtained through the combined actions of an
external magnetic field and an imposed current. It does not require impurity
scattering. Although our model is too crude to reproduce quantitative
aspects of the zero-resistance experiments, we think the possibility of
obtaining a negative conductivity without resorting to impurity scattering
is important.

The elementary geometry of our model allows us to obtain results from
derivations and approximations which are relatively transparent. The model
has some appealing aspects. The negative current is produced by the
excitation of \textit{propagating} states moving against the applied
electric field. Particles in these propagating states can move a relatively
long distance in the ``backward'' direction. This differs from the theories
which invoke backward ``jumps'' with displacements limited by the size of a
Landau orbit.

The backward propagating excited states in our model have their origin in
the reduced symmetry of the band structure. An imposed current gives the
microwave-excited states a preferred direction of motion because the
combination of the magnetic field and the Hall field reduce the band
structure symmetry.

Our model has significant limitations. The lattice approximation means there
is no clearly identified cyclotron resonance, and the one-dimensional
geometry obscures the connection between a negative conductivity and a
zero-resistance state. Also, anomalous conductivity is seen only when an
appreciable magnetic flux threads each unit cell of the lattice. This
implies extremely large fields (or a distance unit which is large compared
to the interatomic spacing). Despite these problems, we believe the
counter-flowing itinerant states produced by the reduced band structure
symmetry may relate to the observed zero-resistance states. A generalization
of the results presented here to a more realistic geometry is clearly a
desirable goal.

Consider then, a system where electrons hop from site to nearest-neighbor
site on the essentially one-dimensional lattice illustrated in Figure 1.
This is a simplification of the famous model described by Hofstadter \cite
{Hof} and many others \cite{Azbel},\cite{Harper}. The restriction to only
two rows of lattice sites allows one to add the Hall effect and describe the
magnetic properties in greater detail.

In order to simplify notation and avoid confusing signs, units are scaled,
and the electrons are replaced with positively charged fermions. Taking the
lattice constant to be unity means the two rows of sites $\left( x,y\right) $
are at $y=$[any integer] and $x$ is either $1$ or $2$. The hopping matrix
element in zero field is set equal to $-1$. The magnetic flux (directed
along $+\hat{z}$) through a unit square in the lattice is $\phi $, and the
flux is scaled so $\phi =\pi $ corresponds to one flux quantum penetrating
each unit cell of the lattice. The Hall effect is included by applying a
potential ($-v$ on the $x=1$ row, and $+v$ on the $x=2$ row).

The magnetic field changes the phases of the hopping matrix elements \cite
{Peierls}. Hopping to larger $y$ (up) on the $x=2$ row and hopping to
smaller $y$ (down) on the $x=1$ row are both multiplied by $\exp (i\phi )$.
(Matrix elements for hopping in opposite directions must be complex
conjugates.) Matrix elements for hopping between the $x=1$ and $x=2$ rows
remain equal to $-1$.

Consider Bloch waves traveling in the $y-$direction ($\psi \approx \exp
(iky) $). For these states, the Schroedinger equation reduces to 
\begin{eqnarray}
\varepsilon (k)\psi _1 &=&-v\psi _1-\psi _2-\exp (-i[k+\phi ])\psi _1-\exp
(i[k+\phi ])\psi _1  \label{Schr} \\
\varepsilon (k)\psi _2 &=&+v\psi _2-\psi _2-\exp (-i[k-\phi ])\psi _2-\exp
(i[k-\phi ])\psi _2  \nonumber
\end{eqnarray}
where $\psi _1$ and $\psi _2$ are the wave amplitudes on the $x=1$ and $x=2$
rows. The resulting two eigenvalues (for each $k$) yield the energy bands. 
\begin{equation}
\varepsilon _{\pm }(k)=-2\cos (k)\cos (\phi )\pm \sqrt{\left( 2\sin (k)\sin
(\phi )-v\right) ^2+1}  \label{Epm}
\end{equation}
Example band structures of this system are shown in Figures 2a,b. In Figure
2a, $\phi =\pi /5$ and $v=0$. The symmetry $\varepsilon (+k)=\varepsilon
(-k) $ follows from the system's two-fold rotational axis. (There is no
time-reversal symmetry in a magnetic field.) Also shown in Figure 2a is the
Fermi energy $E_F$ for a half-filled band and the Fermi wave numbers $k_L$
and $k_R$. When there is no current and $v=0$, $k_L=-k_R$.

Imposing a current in this system generates a transverse Hall voltage which
appears as the potential $v$ in the Schroedinger equation (Eq.(\ref{Schr}))
and in the energy levels (Eq.(\ref{Epm})). Figure 2b shows the reduced
symmetry of the $\phi =\pi /5$ energy bands which results from $v=0.2$. The
reduced symmetry means the right and left Fermi wave numbers ($k_L$ and $k_R$%
) are no longer equal in magnitude.

The current $I$ is produced because particles in the interval $%
k_L<k<k_L+\Delta k$ are moved to the interval $k_R<k<k_R+\Delta k$, as
represented by the open and filled circles in the lower band in Figure 2b.
In scaled units, the current is 
\begin{equation}
I=\frac 1{2\pi }\int_{k_L+\Delta k}^{k_R+\Delta k}\frac{\partial \varepsilon
(k)}{\partial k}dk=\frac 1{2\pi }\left( \varepsilon (k_R+\Delta
k)-\varepsilon (k_L+\Delta k)\right)
\end{equation}

The Hall potential is determined by requiring the total charge on the $x=1$
and $x=2$ rows to be essentially the same. This will minimize the
electrostatic energy. For a given $k$, the wave function amplitudes on the
two rows ($x=1$ and $x=2$ in the lower band) are 
\begin{equation}
\left( 
\begin{array}{l}
\psi _1 \\ 
\psi _2
\end{array}
\right) =\left( 
\begin{array}{l}
\sin (\theta (k)/2) \\ 
\cos (\theta (k)/2)
\end{array}
\right) 
\end{equation}
where 
\begin{equation}
\cos \theta (k)=\frac{2\sin (k)\sin (\phi )-v}{\sqrt{\left( 2\sin (k)\sin
(\phi )-v\right) ^2+1}}  \label{Theta}
\end{equation}
with $0<\theta (k)<\pi $. Note that 
\begin{equation}
\varepsilon _{+}(k)-\varepsilon _{-}(k)=\frac 2{\sin \theta (k)}  \label{DE}
\end{equation}
and the difference between the probability density on the two rows is 
\begin{equation}
\rho _2(k)-\rho _1(k)=\cos \theta (k)  \label{Rho}
\end{equation}
The charge difference $\rho _2(k)-\rho _1(k)$ is positive for $k>0$ (for
small enough $v$). This is the expected sign for the Hall effect. Positive
particles (in the lower band) moving along $+y$ in a $z-$directed magnetic
field will be pushed in the $+x$ direction.

The Hall voltage is determined self-consistently. In a linear approximation,
the difference between the charge on the $x=1$ and $x=2$ rows can be written
as $\rho _2-\rho _1=\alpha I-\beta v$, where $\rho _2$ and $\rho _1$ are the
total charges on the two rows, and $\alpha $ and $\beta $ are
proportionality constants. However, electrostatics mean $v=\gamma \left(
\rho _2-\rho _1\right) $, where $\gamma $ is another constant. Physically,
one expects $\gamma $ to be large because a small charge imbalance leads to
a large electrostatic potential. The linearized relation $\rho _2-\rho
_1=\alpha I/\left( 1+\beta \gamma \right) $ implies a vanishing charge
imbalance in the limit of large $\gamma $. We use this near-neutrality
condition and Eq.(\ref{Rho}) to obtain the Hall potential $v$. 
\begin{equation}
\frac 1{2\pi }\int_{k_L+\Delta k}^{k_R+\Delta k}\cos \theta (k)dk\rightarrow
0  \label{neutral}
\end{equation}
For example, when $\phi =\pi /5$ the Fermi surface displacement of $\Delta
k\cong 0.33$ is required to make $v=0.2$. This case is illustrated in Figure
2b.

The asymmetric form of the energy bands resulting from a nonzero Hall
potential enables the electromagnetic excitation of a current antiparallel
to the applied field. An $ac$ electric field with frequency $\omega $ will
vertically excite electrons from the lower to the upper energy band. For the 
\textit{ac} electric field polarized along $\hat{x}$, the transition
probabilities are proportional to the squared matrix elements 
\begin{equation}
\left| M\right| ^2\approx \left| \left\langle \psi (+)\right| x\left| \psi
(-)\right\rangle \right| ^2\approx \sin ^2\theta (k)
\end{equation}
The perturbation expression for the induced current is proportional to the
radiation intensity multiplied by the integral 
\begin{equation}
A(\hbar \omega )=\int_{k_L+\Delta k}^{k_R+\Delta k}\delta \left( \hbar
\omega -\left( \varepsilon _{+}(k)-\varepsilon _{-}(k)\right) \right) \sin
^2\theta (k)\frac \partial {\partial k}\left( \varepsilon
_{+}(k)-\varepsilon _{-}(k)\right) dk
\end{equation}
The terms in the integrand represent energy conservation, the squared matrix
element, and the difference between the speeds in the upper and lower bands,
respectively. Using the delta-function identity 
\begin{equation}
\int \delta \left( f(x)\right) dx=\frac 1{\left| df/dx\right| _{f=0}}
\end{equation}
and Eq.(\ref{DE}), one concludes that contributions to $A(\hbar \omega )$
will cancel if the energy-conservation condition is met for two values of $k$%
, since $\partial (\varepsilon _{+}(k)-\varepsilon _{-}(k))/\partial k$ will
be positive for one $k$ and negative for the other. For the example shown in
Figure 2b, non-cancelling contributions to $A(\hbar \omega )$ lie in the
restricted range of $k$ (shaded with the large arrow) shown in Figure 2b.
Thus 
\[
A(\hbar \omega )=-\left( \frac 2{\hbar \omega }\right) ^2\left\{ 
\begin{array}{lll}
1 &  & E(-)<\hbar \omega <E(+) \\ 
0 &  & otherwise
\end{array}
\right. 
\]
where 
\[
E(-)=\varepsilon _{+}\left( k_R+\Delta k\right) -\varepsilon _{-}\left(
k_R+\Delta k\right) 
\]
and 
\[
E(+)=\varepsilon _{+}\left( k_L+\Delta k\right) -\varepsilon _{-}\left(
k_L+\Delta k\right) 
\]

We emphasize that the negative microwave-induced current obtained from this
model is a consequence of the combined effects of an external magnetic field
and a Hall voltage produced by the imposed current. 

FIGURES

Figure 1. A diagram of the lattice. The directions of the magnetic field, $B$%
, and current $I$ are show along with the applied Hall voltages, $\pm v$.

Figure 2a. The symmetric energy bands obtained when the scaled magnetic flux
through each unit cell is $\phi =\pi /5$. There is no current, so the Hall
voltage $v$ vanishes. The wave numbers at the Fermi energy $E_F$ satisfy $%
k_L=-$ $k_R$.

Figure 2b. The energy bands of Figure 2a altered by the Hall voltage ($%
v=0.2) $. The emptied states on the left and the extra filled states on the
right (denoted by open and filled circles) give rise to the current.
Electromagnetic excitations in the shaded $k-$space region with the arrow
give rise to an additional current which opposes the applied current.

\end{document}